\documentclass[prr,aps,twocolumn,floatfix,citeautoscript,showpacs]{revtex4-2}
\usepackage{graphicx,rotating,subfigure,amsmath,amsfonts,amssymb,delarray,color}
\usepackage{hyperref}
\usepackage{xcolor}
\usepackage{soul}
\usepackage{dsfont}
\usepackage[T1]{fontenc}
\usepackage{physics}
\usepackage{multirow}
\usepackage{comment}
\renewcommand{\vec}[1]{\boldsymbol #1}

\newcommand{\im}{\text{i}}

\def\12{\frac{1}{2}}

\begin{document}
\bibliographystyle{apsrev}
\title{Operational Entanglement of Symmetry-Protected Topological Edge States}
\author{Kyle Monkman}
\author{Jesko Sirker}
\affiliation{“Department of Physics and Astronomy and Manitoba
Quantum Institute, University of Manitoba, Winnipeg, Canada R3T 2N2}
\date{\today}
\begin{abstract}
We use an entanglement measure that respects the superselection of
particle number to study the non-local properties of
symmetry-protected topological edge states. Considering half-filled
$M$-leg Su-Schrieffer-Heeger (SSH) ladders as an example, we show that
the topological properties and the operational entanglement
extractable from the boundaries are intimately connected. Topological
phases with at least two filled edge states have the potential to
realize genuine, non-bipartite, many-body entanglement which can be
transferred to a quantum register. The entanglement is extractable
when the filled edge states are sufficiently localized on the lattice
sites controlled by the users. We show, furthermore, that the onset of
entanglement between the edges can be inferred from local particle
number spectroscopy alone and present an experimental protocol to
study the breaking of Bell's inequality.
\end{abstract}
\maketitle
\section{Introduction}
Entanglement in a quantum system is an essential resource for quantum
algorithms and for quantum encryption keys. In order to use this
resource, an important question is how much of the entanglement
present in a quantum state is accessible and can be transfered by
local operations to a quantum register consisting of distinguishable
qubits. A fundamental difficulty in addressing this question for
many-body states of itinerant particles is that the entangled entities
itself are indistinguishable. Suppose observers, Alice and Bob, have
access to two {\it spatially-separated} parts of a quantum system of
indistinguishable particles with a conserved particle number N. Then
there is always a local operation that will collapse the pure or mixed
quantum state $\rho_{AB}$ they share into a state
$\rho_{AB}^{n_A,n_B}$ with fixed local particle numbers $n_A$ ($n_B$)
for Alice (Bob). The operational entanglement (also called accessible
entanglement or entanglement of particles) which can be transferred to
a quantum register is thus given by \cite{WisemanVaccaro,DowlingDohertyWiseman}
\begin{equation}
\label{VW}
E_{\textrm{op}} = \sum_{n_A,n_B} p(n_A,n_B) E[\rho_{AB}^{n_A,n_B}] 
\end{equation}
where $p(n_A,n_B)=\tr\rho_{AB}^{n_A,n_B}$ is the probability to
project onto a state with $n_A,n_B$ particles in the two subsystems
and $E[\rho]$ is an entanglement measure applied to the normalized
projected state. The superselection of particle number also gives rise
to an additional non-local resource associated with the particle
number fluctuations
\cite{SchuchVerstraeteCirac,SchuchVerstraeteCirac2}. To characterize this second, 
complementary resource we introduce the {\it generalized number
entropy} (Shannon entropy)
\begin{equation}
\label{EN}
E_n = -\sum_{n_A,n_B} p(n_A,n_B) \ln  p(n_A,n_B) \; .
\end{equation}
We note that $E_n=0$ if $n_A$ and $n_B$ are fixed, $E_n$ has an upper
bound
$E^{\textrm{max}}_n=\ln[(n_A^{\textrm{max}}+1)(n_B^{\textrm{max}}+1)]$
if there are at most $n_A^{\textrm{max}}$ ($n_B^{\textrm{max}}$)
particles in $A$ ($B$), and $E_n$ does not increase under local
operations.

For a bipartition of a pure state $\rho$, one can use the von-Neumann
entanglement entropy $E_{vN}$ as the entanglement measure. In this
case, $E_{\textrm{vN}}=E_n + E_{\textrm{op}}$ where the operational or
configurational entropy is now given by Eq.~\eqref{VW} with
$E[\rho]\equiv E_{\textrm{vN}}[\rho]$
\cite{KlichLevitov,Rakovszky2019,Bonsignori2019,MurcianodiGiulio,MurcianodiGiulio2,BarghathiCasianoDiaz} 
and the restriction $n_B=N-n_A$ is placed on the sums in
Eqs.~(\ref{VW}, \ref{EN}). The number entropy for a bipartition has
recently been measured in a cold atomic gas experiment
\cite{LukinGreinerMBL} and can be used to obtain a bound on the total 
entanglement entropy \cite{KieferUnanyan1,KieferUnanyan2}.

\begin{figure}[!t]
\includegraphics[width=0.99\columnwidth]{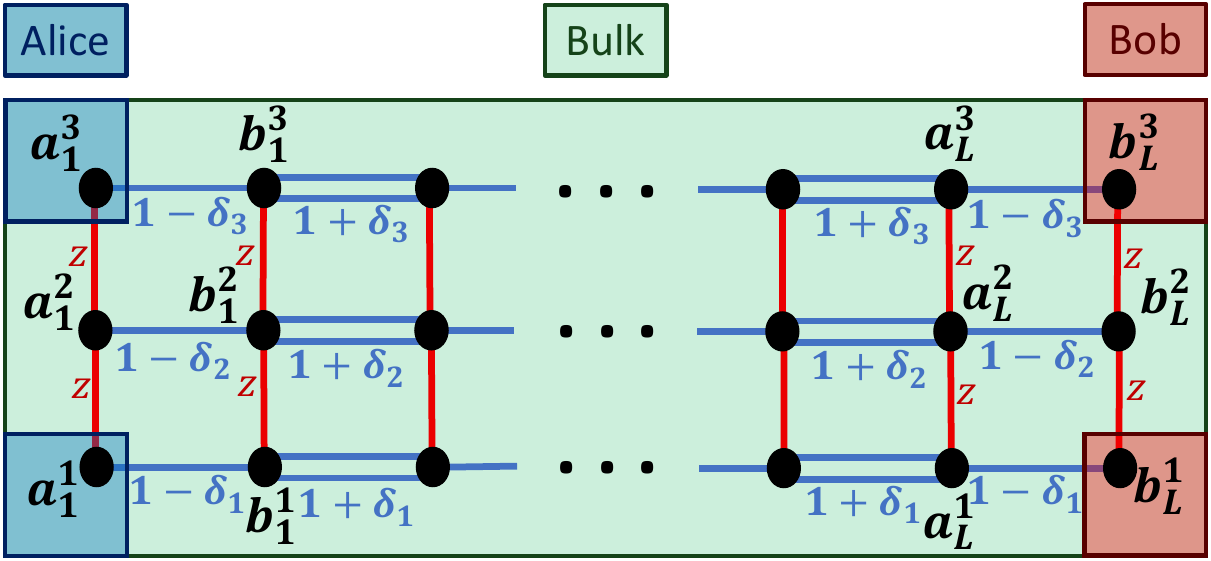}
\caption{$3$-leg SSH ladder of spinless fermions. 
Along the chain direction the hopping amplitude alternates between
$1-\delta_s$ and $1+\delta_s$ while $z$ is the intrachain
coupling. Alice and Bob only control a few sites at opposite edges of
the ladder.}
\label{Fig1}
\end{figure}
Long-range entanglement in many-body systems is, in general, very
fragile against local perturbations. An exception are edge states in
systems with a topologically non-trivial bulk. In this case, the
number of edge states is topologically protected and cannot be changed
without closing the gap or breaking the symmetry
\cite{RyuSchnyder,RyuSchnyderReview}. Since topological edge states
can exist simultaneously on multiple edges, a filled edge state is a
non-local quantum resource.

In this article we will show that symmetry-protected topological edge
states in lattice models can be a source of genuine,
spatially-separated, {\it non-bipartite} many-body entanglement which
can be transfered to a quantum register. We will show, furthermore,
that the onset of entanglement between the edges can be inferred from
local particle number spectroscopy, making it measurable, for example,
by current-day cold atomic gas experiments
\cite{LukinGreinerMBL,BakrGillen,ShersonWeitenberg}. We also present an 
experimental protocol to test Bell's inequalities
\cite{BellInequality,BellReview,CHSH,HarocheBook,CHSH-exp}.
We find that for half-filled pure states of arbitrary length, two or
more filled edge states are required to break the inequality.
Furthermore, using the grand canonical and canonical ensemble, we
establish the {\it spatially separated}, operational breaking of
Bell's inequality of a many-body system in thermal equilibrium.

In the following, we will consider lattice models of itinerant
spinless fermions. Let us first briefly discuss why in this case a
single edge state is insufficient to produce any operational
entanglement. Let us assume that the edge state is very sharply
localized at the left and right edges of a chain which are controlled
by Alice and Bob, respectively. To simplify the argument, we ignore
the bulk completely for now and assume that the edge state is a
single-particle pure state. A maximally entangled edge state in
occupation number representation is then of the form $|\Psi\rangle =
(|10\rangle \pm |01\rangle)/\sqrt{2}$ and has von-Neumann entanglement
entropy $E_{\textrm{vN}}=\ln 2$. If Alice and Bob measure their local particle
numbers in order to perform local operations, this state however
collapses onto the product state $|\Psi^{1,0}\rangle
\propto |10\rangle$ or $|\Psi^{0,1}\rangle \propto |01\rangle$. 
Thus, this state has only number entropy $E_n$ but no operational
entanglement. We therefore need at least two filled edge states to
have any operational entanglement. In the latter case, the projected
state $|\Psi^{1,1}\rangle$ can have operational entanglement while the
states $|\Psi^{2,0}\rangle$ and $|\Psi^{0,2}\rangle$ are again product
states.
\section{Model}
In order to build a system with more than one filled edge state at
half-filling, we couple Su-Schrieffer-Heeger (SSH) chains
\cite{SSH,SuSchriefferHeegerRMP} to form $M$-leg ladders. While there
are many other possible choices, the SSH chain is one of the simplest
systems with non-local, symmetry-protected topological edge
states. Furthermore, its properties can be studied experimentally
using cold atoms in optical superlattices
\cite{AtalaAidelsburger,XieGou,MeierAn}. 
Open boundary conditions can be implemented using an optical box
potential \cite{Meyrath,Gaunt}. As we will show below, the operational
entanglement that results from the topological edge states can be
observed in this system using modern experimental techniques.

Let $J$, $\delta_s$ and $z$ be real constants that indicate hopping between
sites, and ${a_j^s}^\dag$ (${b_j^s}^\dag$) creation operators of an `a'
(`b') spinless fermion on chain $s$, in unit cell $j$. The unit cell
consists of $2M$ elements, two elements on each of the $M$ chains. The
Hamiltonian of the model in second quantization is then given by
\begin{eqnarray}
\label{Ham}
H &=& J \sum_{j,s} \lbrace (1-\delta_s){a_{j}^s}^{\dag}b_{j}^s + (1+\delta_s){b_{j}^s}^{\dag}a_{j+1}^s + h.c. \rbrace \nonumber \\
&+& z \sum_{j,s}  \lbrace {a_{j}^s}^{\dag} a_{j}^{s+1}  + 
{b_{j}^s}^{\dag} b_{j}^{s+1} + h.c. \rbrace
\end{eqnarray}
with $J=1$ and $|\delta_s| \leq 1$ for the remainder of the paper. A visualization of the SSH ladder 
for $M=3$ with open boundary conditions is shown in Fig.~\ref{Fig1}.

The non-interacting SSH ladder, a member of the BDI symmetry class, has
three non-spatial symmetries. These symmetries are time reversal
$\hat{T}$, charge-conjugation $\hat{C}$, and chiral $\hat{S}$, see
Ref.~\cite{RyuSchnyderReview}. The relevant symmetry here is the
chiral symmetry. $\hat{S}$ is defined in terms of its action on
annihilation operators as
\begin{equation}
\label{symm}
\hat{S} a_j^s \hat{S}^{-1} = (-1)^{s+1} {a_j^s}^\dag , \  
\hat{S} b_j^s \hat{S}^{-1} = (-1)^s{b_j^s}^\dag ,
\end{equation}
with $\hat{S} i \hat{S}^{-1} = -i$. The Hamiltonian in Eq.~\eqref{Ham}
satisfies the symmetries $\hat{T} H
\hat{T}^{-1} = H$, $\hat{C} H \hat{C}^{-1} = H$, and $\hat{S} H \hat{S}^{-1} = H$ 
for any values of $J$, $\delta_s$ and $z$. There are also additional
non-spatial symmetries present when certain restrictions are placed on
the parameters. In particular, additional chiral symmetries enable
additional topological invariants
\cite{WakatsukiEzawaTanakaNagaosa} which are discussed further in App.~C. 
\begin{figure}[!ht]
\includegraphics[width=0.99\columnwidth]{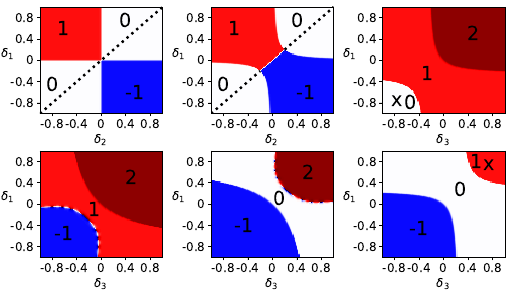}
\caption{Topological phase diagram for $M=2$ and $M=3$ leg ladders. 
The numbers indicate the value of the topological invariant $I$. A
dotted line or an 'x' is placed where the analytical result,
Eq.~\eqref{winding_ana}, applies. Top, left to right: $M=2$, $z=0.0$;
$M=2$, $z=0.4$; $M=3$, $z=0.9$, $\delta_2=-0.75$. Bottom, left to
right: $M=3$, $z=0.9$, $\delta_2=-0.25$; $M=3$, $z=0.9$,
$\delta_2=0.25$; $M=3$, $z=0.9$, $\delta_2=0.75$.}
\label{Fig2}
\end{figure}
\section{Topology}
In order to find the number of edge states of the $M$-leg SSH ladder
for a given set of parameters, we have to analyze its topological
phase diagram. To define the winding number $I$ for our system with
chiral symmetry, we follow Ref.~\cite{Gurarie}.

First, we define a unitary matrix $U_S$ for the chiral symmetry
$\hat{S}$ from the condition $\hat{S} \psi_n \hat{S}^{-1} = \sum_{m}
(U_S)_{n,m}^* \psi_{m}^\dag$, where $\psi_n$ are the $a_j^s$ and
$b_j^s$ operators in an arbitrary basis. $U_S$ has the property
$\mbox{tr}\, U_S=0$ and we define the phase such that
$U_S^2=\mathbb{I}$. $U_S$ can be put into a momentum space, block
diagonal form represented by $U_S(k)$.  Let $g(k)=H^{-1}(k)$ be the
matrix representation of the single particle Green's function
corresponding to the Hamiltonian $H$ in Eq.~\eqref{Ham}. Then the
topological invariant is given by
\cite{Gurarie,WakatsukiEzawaTanakaNagaosa,Padavic}
\begin{equation}
\label{winding}
I = \frac{1}{4 \pi \im } \mbox{tr} \int dk \ {U_S}(k) g^{-1}(k) \partial_k g(k),
\end{equation}
which, for non-interacting systems, is equivalent to the winding
number as defined, for example, in Ref.~\cite{RyuSchnyderReview}. We
prove the equivalence of the invariants in App.~A. We note,
furthermore, that the related Zak phase for a single SSH chain has
been measured experimentally in cold atomic gases
\cite{AtalaAidelsburger}. Before analyzing the full topological phase
diagram for $M=2$ and $M=3$ leg ladders numerically, we first note the
following important analytical result for the general $M$-leg case: Suppose
that $\delta_1=\delta_2 = \dots \equiv \delta$ and $0 \leq
|z|\cos(\frac{\pi}{M+1})< |\delta|$. (i) If $\delta < 0$, then
$I=0$. (ii) If $\delta > 0$, then
\begin{equation}
\label{winding_ana}
I=1 \quad \mbox{for $M$ odd,} \;\; \mbox{and} \quad I=0 \quad \mbox{for $M$ even.} 
\end{equation} 
The proof of this result can be found in App.~B. We note that this
even-odd effect resembles the celebrated result for spin-$1/2$ ladders
\cite{DagottoRice}.

The topological phase diagram based on the invariant defined in
Eq.~\eqref{winding} is presented in Fig.~\ref{Fig2} for ladders with
$M=2$ and $M=3$. A dotted line or an 'x' is placed where the
analytical result, Eq.~\eqref{winding_ana}, applies. A similar phase
diagram for the $M=2$ case using a different parameterization can be
found in Ref.~\cite{Padavic}. When $I=0$ we have no edge states and
when $I =
\pm 1$ we have a single filled edge state. The case which we want 
to concentrate on in the following is $I=2$ where we have two filled
edge states.

\section{Entanglement}
The ground state of the SSH ladder is described by a pure state wave
function $\rho = |\Psi \rangle \langle \Psi|$. We imagine however,
that Alice and Bob have access to only a small number of sites at the
edges of the system. Note that in contrast to the often studied case
of a bipartition, tracing out the rest of the system leads to a
{\it mixed state} $\rho_{AB}$ which is our starting point.

Now that we have a density matrix $\rho_{AB}$ that only describes the
two subsystems we are interested in, we can apply the operational
entanglement measure defined in Eq.~\eqref{VW} with $E[\rho]$ being a
bipartite measure of entanglement for a mixed state. Regardless of the
chosen mixed state measure of entanglement, $E_{\textrm{op}}$ is not
easy to compute in general. However, for small dimensions of
$\rho_{AB}$ a calculation of its matrix elements using correlation
functions is feasible
\cite{DowlingDohertyWiseman}. For the case of two edge states ($I=2$) 
considered here the only projected density matrix which will
contribute to $E_{\textrm{op}}$ is $\rho_{AB}^{1,1}$. We will call the
two modes on one side of the ladder $A_1$ and $A_2$, and the modes on
the other side $B_1$ and $B_2$. Next, we define the projection
operators $P^A = A_1^\dag A_1(1-A_2^\dag A_2)+ (1-A_1^\dag
A_1)A_2^\dag A_2$ and analoguously $P^B$ which project the ground
state onto a (non-normalized) state with a single particle in each
subsystem, $|\Psi^{1,1}\rangle = P^A P^B |\Psi\rangle$. The matrix
elements of the $4 \times 4$ matrix $\rho_{AB}^{1,1}$ can now be
computed from correlation functions in the projected state,
\begin{eqnarray}
\label{rho11}
(\rho_{AB}^{1,1})_{i i'}^{j j'} &=& \tr_{A,B}(\rho_{AB}^{1,1} A_j^\dag B_{j'}^\dag {B_{i'}} A_i) \nonumber \\ 
&=& \langle \Psi^{1,1}|A_j^\dag B_{j'}^\dag {B_{i'}} A_i|\Psi^{1,1} \rangle. 
\end{eqnarray}
A more detailed description of how to calculate the matrix elements of
$\rho_{AB}^{1,1}$ is given in App.~D. In systems---such as the SSH
ladder considered here---which are Gaussian, we can use Wick's Theorem
to turn the multi-point correlation functions into products of
two-point correlation functions. Since we already know from the
topological phase diagram, Fig.~\ref{Fig2}, that we need at least a
$3$-leg ladder to have two filled edge states, we concentrate on this
case in the following. The choice of the few sites controlled by Alice
and Bob needs to be based on prior knowledge of where the topological
edge states are primarily located. Based on numerical results for
strong dimerizations, we choose $A_1 = a_1^1$, $A_2 = a_1^3$, $B_1 =
b_L^1$ and $B_2 = b_L^3$, see Fig.~\ref{Fig1}.
\begin{figure}
\centering
\includegraphics[width=0.99\columnwidth]{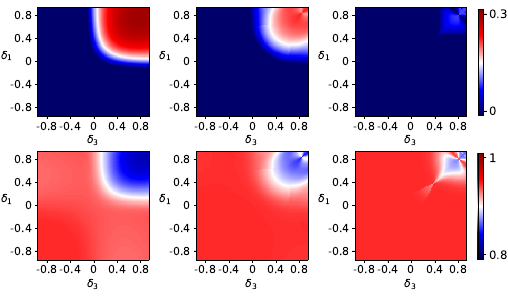}
\caption{Entanglement phase diagram based on $E_{\textrm{op}}=p(1,1)E_\textrm{neg}[\rho_{AB}^{1,1}]$ 
in units of $\ln 2$ and generalized number entropy $E_n$ in units of
$\ln 9$ for $3$-leg ladders with $L=16$. Results for $E_\textrm{op}$
are shown in the top panels and for $E_n$ in the bottom panels. Left
to right: $z=0.9$, $\delta_2=-0.75$; $z=0.9$, $\delta_2=-0.25$;
$z=0.9$, $\delta_2=0.25$.}
\label{Fig3}
\end{figure}
There are many different entanglement measures one can use to quantify
the entanglement of a mixed-state density matrix
\cite{Horodecki}. Here we use an additive measure of entanglement, the
logarithmic negativity
\cite{Zyczkowski,VidalWerner},
\begin{equation}
E_{\textrm{neg}}[\rho_{AB}] = \ln(|| \rho_{AB}^{T_A}||)\, ,
\end{equation}
where $\rho_{AB}^{T_A}$ is the partial transpose with respect to Alice
and $|| \rho_{AB}^{T_A} ||$ is the trace norm of the normalized
matrix. Results for
$E_\textrm{op}=p(1,1)E_\textrm{neg}[\rho_{AB}^{1,1}]$ for the $3$-leg
ladder---using the same dimerizations $\delta_s$ as in the topological
phase diagrams---are shown in Fig.~\ref{Fig3}.  For $\delta_2=-0.75$
and $\delta_2=-0.25$, the regions where
$E_{\textrm{neg}}[\rho_{AB}^{1,1}]\approx \ln 2$ coincide with the
regions in Fig.~\ref{Fig2} with winding number $I=2$. The topology of
the system is thus directly tied to the operational entanglement which
can be extracted from the system by Alice and Bob. However for
$\delta_2=0.25$, the region with operational entanglement is much
smaller than the region with $I=2$. In this case, the edge states are
not sufficiently localized. Alice and Bob would need to control more
sites to extract all of the operational entanglement present in the
two edge states. Virtually identical results are obtained if we use
the entanglement of formation instead of the logarithmic negativity,
see App.~E. From an experimental perspective, the most important
result however is that the regions with operational entanglement and
winding number $I=2$ can be identified by simply measuring the
generalized number entropy $E_n$, Eq.~\eqref{EN}, on a small number of
sites only, see bottom panels in Fig.~\ref{Fig3}. When the edge states
are forming, $p(n_A=1,n_B=1)$ increases leading to a decrease of
$E_n$. The number entropy can be measured straightforwardly in cold
atomic gases by single-site atomic imaging as has recently been
demonstrated in Ref.~\cite{LukinGreinerMBL}.
\begin{figure}[!t]
\includegraphics[width=0.99\columnwidth]{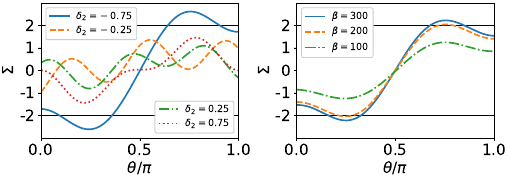}
\caption{$\Sigma$, as defined in Eq.~\eqref{Sigma}, for a $3$-leg ladder. 
Left panel: $T=0$ with $L=16$, $\delta_1=0.9$, $\delta_3=0.8$, and
$z=0.9$ for different $\delta_2$. For $\delta_2=-0.75$, corresponding
to a winding number $I=2$, the CHSH inequality is broken. Right panel:
$L=8$, $\delta_1 = 0.29$, $\delta_2 = -0.99$, $\delta_3 = 0.99$, $z
= 0.95$ and chemical potential $\mu=0$ for various inverse temperatures $\beta=1/T$.}
\label{Fig4}
\end{figure}
\section{Bell's theorem}
Moving beyond the indirect observation of the entanglement between the
edges by monitoring the number entropy, one of the most fundamental
ways to prove that two qubits are entangled is to show that a Bell
inequality is broken. Here we will choose the Clauser, Horne, Shimony
and Holt (CHSH)
\cite{CHSH,CHSH-exp} version of Bell's inequality. Let $\vec{a}$, $\vec{a}'$,
$\vec{b}$, and $\vec{b}'$ be three-dimensional vectors. Let
$\vec{\sigma}^{A/B}$ be the vector of $A/B$ Pauli matrices. Defining
$\langle \dots \rangle_{1,1} = tr(\rho_{AB}^{1,1}\dots )/p(1,1)$, the
CHSH inequality reads
\begin{eqnarray}
\label{Sigma}
&& -2 \leq \Sigma_{\vec{a}, \vec{a}', \vec{b}, \vec{b}'}  \leq 2\, ,  \qquad \mbox{with} \\
&&\Sigma_{\vec{a}, \vec{a}', \vec{b}, \vec{b}'} = \langle (\vec{a} \cdot \vec{\sigma}^A) (\vec{b} \cdot \vec{\sigma}^B) \rangle_{1,1} - \langle(\vec{a}' \cdot \vec{\sigma}^A) (\vec{b} \cdot \vec{\sigma}^B) \rangle_{1,1} \nonumber \\ 
&& + \langle(\vec{a} \cdot \vec{\sigma}^A) (\vec{b}' \cdot \vec{\sigma}^B) \rangle_{1,1} + \langle(\vec{a}' \cdot \vec{\sigma}^A) (\vec{b}' \cdot \vec{\sigma}^B) \rangle_{1,1}. \nonumber
\end{eqnarray}
To show the breaking of the inequality, we choose the vectors
$\vec{a}$, $\vec{a}'$, $\vec{b}$ and $\vec{b}'$ to be in the x-z plane
with $\vec{a} \cdot \vec{\sigma}^A = \cos \theta_a \sigma_z^A+\sin
\theta_a \sigma_x^A$ and $\vec{b} \cdot \vec{\sigma}^B = \cos \theta_b
\sigma_z^B+\sin \theta_b \sigma_x^B$. We use the representation of the
Pauli operators $\sigma_x^A = A_1^\dag A_2 + A_2^\dag A_1$,
$\sigma_y^A = -i A_1^\dag A_2 + i A_2^\dag A_1$, $\sigma_z^A =
A_1^\dag A_1 - A_2^\dag A_2$ and similarly for
$\sigma_{x,y,z}^B$. Results for the $3$-leg ladder are shown in
Fig.~\ref{Fig4} with $\theta\equiv\theta_a=\theta_{b'}/2=\theta_{a'}/3$ and
$\theta_b=0$. 
For $\delta_2=-0.75$, corresponding to a winding number $I=2$ (see top
right panel in Fig.~\ref{Fig2}), the CHSH inequality in the projected
state $\rho_{AB}^{1,1}$ is broken. Note that while $\delta_2=-0.25$
and $\delta_2=0.25$ also correspond to $I=2$, the edge states are not
sufficiently localized in these cases to break the CHSH inequality. 

We can also evaluate the correlators in the system in thermal
equilibrium for the grand canonical and canonical ensembles. At zero
temperature, there are two filled edge states and two empty edge
states. The energy gap between the filled and empty edge states
decreases as the system size is increased. At finite temperature, this
energy gap is directly related to the filling of the four edge
states. In the grand canonical case, we require the chemical potential
$\mu=0$ in order to maintain particle-hole symmetry. The right panel
of Fig.~\ref{Fig4} shows that for the grand canonical ensemble, the
CHSH inequality is broken for temperatures up to $\beta = 1/T \approx
200$. The canonical case is discussed in App.~F.
\begin{figure}[!t]
\includegraphics[width=0.99\columnwidth]{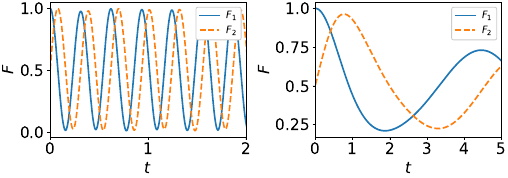}
\caption{Fidelities \eqref{fidelity} showing that the 
time evolution with the Hamiltonian \eqref{Hprime} leads to an
effective rotation of the density matrix by $\pi/2$ around the
$y$-axis. When $\kappa=10$ (left), there is a nearly perfect
rotation. $\kappa=1$ (right) does not result in a purely rotational
process. Both figures have parameters $M=3$, $L=16$, $\delta_1 = 0.9$,
$\delta_2 = -0.75$, $\delta_3 = 0.8$ and $z = 0.9$.}
\label{Fig5}
\end{figure}
\section{Experimental protocol}
Next, we discuss a possible experimental protocol for showing that
Bell's inequality is broken. We can relate elements $\langle \sigma_A^{x,z}
\sigma_B^{x,z} \rangle_{1,1}$ to two-particle correlators of the full
many-body state $|\psi \rangle$. Calculating $\langle \sigma_A^z
\sigma_B^z \rangle_{1,1}$ then amounts to calculating density-density correlations
which are experimentally accessible \cite{Foelling}. A function such
as $\langle \sigma_A^z
\sigma_B^x \rangle_{1,1}$, on the other hand, is more difficult to obtain
experimentally because it involves measuring correlators
such as $\langle A_1^\dag A_1 B_1^\dag B_2 \rangle$, see App.~D.

While for non-interacting systems measuring the single-particle
correlation functions might be possible and is sufficient, we can more
generally make use of the matrix operation $\sigma^z \otimes \sigma^x = (I
\otimes R) (\sigma^z \otimes \sigma^z) (I \otimes R^\dag)$, where $R$ is a
$\pi/2$ rotation matrix about the $y$-axis and $I$ the identity
matrix. In the following, we show that by time evolving the {\it full
many-body state} $|\psi \rangle$, we can implement a rotation operator
$R$ on the two-site density matrix $\rho_{AB}^{1,1}$. To do so, we
use the time evolution operator $\exp(-\text{i}H't)$ with
\begin{equation}
\label{Hprime}
H' = H + \lambda B_1^\dag B_2 + \lambda^* B_2^\dag B_1,
\end{equation}
$H$ defined in Eq.~\eqref{Ham}, and $\lambda$ a constant. We now
compare the two-site rotated density matrix $(I
\otimes R) \rho_{AB}^{1,1} (I \otimes R^\dag)$ with the two-site 
density matrix $\sigma_{AB}^{1,1}(t)$ obtained from the full
time-evolved state $|\Psi(t)\rangle$ using the fidelity function for
density matrices \cite{Bures,Uhlmann,ZanardiQuan,CamposVenuti2011}
\begin{equation}
\label{fidelity}
F(\rho,\sigma) = \left[ \ \tr\sqrt{\sqrt{\sigma}\rho \sqrt{\sigma}} \ \right]^2.
\end{equation}
Since we want to rotate around the $y$-axis, we set $\lambda=-i\kappa$
with $\kappa$ being a real number. We define $F_1(t) =
F(\rho_{AB}^{1,1},\sigma_{AB}^{1,1})$ and $F_2(t) = F((I \otimes R)
\rho_{AB}^{1,1} (I \otimes R^\dag),\sigma_{AB}^{1,1})$ with $F_1(0)=1$
and $F_2(0)=0.5$. The time dependence of $F_1$ and $F_2$ is shown in
Fig.~\ref{Fig5}. For large $\kappa \sim 10$, $F_1$ and $F_2$ oscillate
out of phase with a maximum fidelity close to $1$. This shows that
implementing the coupling \eqref{Hprime} allows for an effective
rotation of the density matrix.
\section{Conclusions}
We have shown that symmetry-protected topological edge states in a
system of itinerant particles can be a resource of spatially
separated, non-bipartite, operational entanglement which can be
transfered to a quantum register of distinguishable qubits. Two edge
states which are both filled and sufficiently localized are required
to obtain two entangled qubits. While we have used an explicit
construction of such edge states based on coupled SSH chains, the
connection established here between the topology of the system and the
amount of entanglement which can be extracted from its edge states is
expected to be general. We have shown, furthermore, that the number
entropy measured on a few sites only is an indirect probe of the
topological and entanglement properties of a system, which is easily
accessible in cold atomic gas experiments. Going one step further, we
have also demonstrated that the non-bipartite operational entanglement
obtained from the projected ground state of the many-body system is
sufficient to break Bell's inequality and presented a protocol to
measure these strong, {\it spatially separated} quantum correlations
experimentally.

\acknowledgments
The authors acknowledge support by the Natural Sciences and
Engineering Research Council (NSERC, Canada) and by the Deutsche
Forschungsgemeinschaft (DFG) via Research Unit FOR
2316. K.M. acknowledges support by the Vanier Canada Graduate
Scholarships Program. We are grateful for the computing resources and
support provided by Compute Canada and Westgrid.

\appendix
\renewcommand{\theequation}{\thesection.\arabic{equation}}

\label{AppA}
\setcounter{equation}{0}
\renewcommand{\thesection}{\text{A}}
\section{Equivalence of Topological Invariants}
We will manipulate 
\begin{equation}
\label{Invariant}
I = \frac{1}{4 \pi \im } \mbox{tr} \int dk \ {U_S}(k) g^{-1}(k) \partial_k g(k)
\end{equation}
with $g(k)=H^{-1}(k)$ into an equivalent form. We first note that
$U_S$ can be put into block diagonal form in momentum space with the
blocks represented by $U_S(k)$. The property $U_S^2=\mathbb{I}$ for
the full matrix implies a similar property of the blocks
$U_S^2(k)=\mathbb{I}_{2M}$ where $M$ is the number of legs of the SSH
ladder. Also since $\hat{S}$ is a non-spatial symmetry,
$\text{tr}\, U_s=0$ implies that $\text{tr}\, U_s(k)=0$ for the individual
blocks as well. Then $U_S^2(k)=\mathbb{I}_{2M}$ and
$\text{tr}\, U_S(k)=0$ imply that we can pick a basis such that
\begin{equation}
\label{BlockUnitary}
U_S(k)= \begin{pmatrix} \mathbb{I}_M & 0 \\ 0 & -\mathbb{I}_M \end{pmatrix}.
\end{equation}
The operator condition $\hat{S} H \hat{S}^{-1}=H$ implies for the momentum blocks 
that $U_S^\dag(k) H(k) U_S(k)=-H(k)$. This condition implies that in the same basis 
as \eqref{BlockUnitary},
\begin{equation}
\label{BlockHamiltonian}
H(k) = \begin{pmatrix} 0 & D(k) \\ D^\dag(k) & 0 \end{pmatrix}.
\end{equation}
Then plugging \eqref{BlockUnitary} and \eqref{BlockHamiltonian} into \eqref{Invariant}, we get
\begin{equation}
\label{Invariant2}
I = \frac{1}{4 \pi i } tr \int dk \  \lbrace D \partial_k D^{-1} - D^\dag \partial_k {D^\dag}^{-1} \rbrace.
\end{equation}
Now, we use the polar decomposition $D(k) = |D(k)| q(k)$, where $|D(k)|$ is positive 
definite and $q(k)$ is unitary. We obtain
\begin{equation}
\label{Invariant3}
I = \frac{\im}{2 \pi} \mbox{tr} \int dk \  q^\dag(k) \partial_k q(k) .
\end{equation}

Now we demonstrate that this method of finding the topological
invariant is equivalent to the method based on projection operators
\cite{RyuSchnyder,RyuSchnyderReview}. The formalism starts by writing
the Hamiltonian $H(k)$ in the off block diagonal basis
Eq.~\eqref{BlockHamiltonian}. Next, we find the column eigenvectors
$v_n(k)$ of Eq.~\eqref{BlockHamiltonian}. We define
\begin{equation}
\label{Pk}
P(k) = \sum_n v_n(k) v_n^\dag(k)
\end{equation}
where the sum is over all eigenvectors $v_n(k)$ with negative eigenvalues. We also define
\begin{equation}
\label{Qk}
Q(k) = \mathbb{I} - 2P(k).
\end{equation}
$Q(k)$ turns out to always be of the form 
\begin{equation}
\label{Qk2}
Q(k) = \begin{pmatrix} 0 & \tilde{q}(k) \\ \tilde{q}^\dag(k) & 0 \end{pmatrix}.
\end{equation}
Then the invariant is calculated by plugging $\tilde{q}(k)$ into Eq.~\eqref{Invariant3} as $q(k)$. 

Now, we only need to prove that $\tilde{q}(k)=q(k)$. To do so, first define
$u_n(k)$ as the normalized eigenvectors of $|D(k)|$. Then the
eigenvectors $v_n(k)$ with negative eigenvalues are
\begin{equation}
\label{vn}
v_n(k) = \frac{1}{\sqrt{2}} \begin{pmatrix} u_n(k) \\ -q^\dag(k) u_n(k) \end{pmatrix}.
\end{equation}
Next, we use Eq.~\eqref{vn} as the vectors in Eq.~\eqref{Pk} to obtain
the projection operators $P(k)$ and $Q(k)$. We will also use the fact
that the $u_n(k)$ vectors form a complete basis. We obtain
\begin{equation}
\label{Qk3}
Q(k) = \begin{pmatrix} 0 & q(k) \\ q^\dag(k) & 0 \end{pmatrix}.
\end{equation}
Hence $\tilde{q}(k)=q(k)$.

\label{AppB}
\setcounter{equation}{0}
\renewcommand{\thesection}{\text{B}}
\section{Proof of the Analytic Result}
\textbf{\textit{Analytic Result.}} Suppose
that $\delta_1=\delta_2 = \dots \equiv \delta$ and $0 \leq
|z|\cos(\frac{\pi}{M+1})< |\delta|$. (i) If $\delta < 0$, then
$I=0$. (ii) If $\delta > 0$, then
\begin{equation}
\label{winding_ana_app}
I=1 \quad \mbox{for $M$ odd,} \;\; \mbox{and} \quad I=0 \quad \mbox{for $M$ even.} 
\end{equation}
\textbf{\textit{Proof.}}
Define for all chains s
\begin{equation}
\label{Fourier}
a_j^s = \dfrac{1}{\sqrt{L}} \sum_{n=1}^{L} e^{ik_n j} a_{k_n}^s \ , \ b_j^s = \dfrac{1}{\sqrt{L}} \sum_{n=1}^{L} e^{ik_n j} b_{k_n}^s
\end{equation}
where $k_n=2\pi n/L$ (we will denote $k_n$ as just $k$ going forward). If $M$ is odd, define 
$\hat{\psi}(k) = \begin{pmatrix} a_k^1 & b_k^2 & a_k^3 & ... & a_k^M & b_k^1 & a_k^2 & ... & b_k^M \end{pmatrix}^T$ 
and if $M$ is even, define 
$\hat{\psi}(k) = \begin{pmatrix} a_k^1 & b_k^2 & a_k^3 & ... & b_k^M & b_k^1 & a_k^2 & ... & a_k^M \end{pmatrix}^T$. 
In either case, the Hamiltonian operator $H$ can be written in terms of $2M\times2M$ block matrices $H(k)$ as
\begin{equation}
\label{}
H = \sum_k \hat{\psi}^\dag(k) H(k) \hat{\psi}(k).
\end{equation}
$H(k)$ takes the form
\begin{equation}
\label{Hk}
H(k) = \begin{pmatrix} 0_M & D(k) \\ D^\dag(k) & 0_M \end{pmatrix}.
\end{equation}
where the $D(k)$ blocks are $M\times M$ matrices. If we define 
$x(k) = |x(k)| e^{i\theta(k)} = (1-\delta)+(1+\delta)e^{-ik}$, $D(k)$ can be written as
\begin{equation}
\label{}
D(k) = 
\begin{pmatrix}
x(k) & z & 0 & \dots \\
z & x^*(k) & z & \dots \\
0 & z & x(k) & \dots \\
\vdots & \vdots & \vdots & \ddots &
\end{pmatrix}.
\end{equation}
Since the symmetry $\hat{S}$ is a non-spatial chiral symmetry, the
symmetry operator in Fourier space can be found by replacing $j
\rightarrow k_n$ in Eq.~(4) of the main text. Then one applies the
condition $\hat{S} \psi_n \hat{S}^{-1} = \sum_{m} (U^S)_{n,m}^*
\psi_{m}^\dag$, where $\psi_n$ are the $a_{k_n}^s$ and $b_{k_n}^s$
operators in an arbitrary basis. In the basis of Eq.~\eqref{Hk},
$U_S(k)$ takes the form
\begin{equation}
\label{}
U_S(k) = 
\begin{pmatrix}
\mathbb{I}_M & 0_M \\
0_M & -\mathbb{I}_M
\end{pmatrix}.
\end{equation}

Define $\hat{e}_1$, $\hat{e}_2$, \dots $\hat{e}_{2M}$ as the unit column vectors in the 
same basis as \eqref{Hk}. For $s=1,2,\dots,M$, define
\begin{equation}
\label{v}
v_s^- = \frac{1}{\sqrt{2}}(-C_s \hat{e}_s + \hat{e}_{s+M})
\end{equation}
\begin{equation}
\label{Eq27}
v_s^+ = \frac{1}{\sqrt{2}}(C_s \hat{e}_s + \hat{e}_{s+M}).
\end{equation}
with
\begin{equation}
\label{Eq28}
C_s = 
\begin{cases}
e^{i\theta} & \text{s odd} \\
e^{-i\theta} & \text{s even}.
\end{cases}
\end{equation}
Then $v_s^\pm$ are the eigenvectors of $H(k)$ when $z=0$. The eigenvalues are 
$E_s^\pm(k) = \pm |x(k)|$. Now we transform $H(k)$ into the $v_s^{\pm}$ basis. 
Let V be defined by the unit column vectors $v_s^{\pm}$ as
\begin{gather}
\label{Eq29}
V = 
\begin{pmatrix}
v_1^{-} & v_2^{-} & \dots & v_M^{-} & v_1^{+} & v_2^{+} & \dots & v_M^{+}
\end{pmatrix}.
\end{gather}
Then 
\begin{gather}
\label{}
V^\dag H(k) V = 
\begin{pmatrix}
H_{1,1}^V(k) & 0_M \\
0_M & H_{2,2}^V(k)
\end{pmatrix}
\end{gather}
where $H_{1,1}^V(k)$ and $H_{2,2}^V(k)$ are $M \times M$ tridiagonal matrices defined by
\begin{gather}
\label{Eq31}
H_{2,2}^V = -H_{1,1}^V = 
\begin{pmatrix}
|x(k)| & z e^{-i\theta(k)} & 0 & \dots \\
z e^{i\theta(k)} & |x(k)| & z e^{i\theta(k)} & \dots \\
0 & z e^{-i\theta(k)} & |x(k)| & \dots \\
\vdots & \vdots & \vdots & \ddots &
\end{pmatrix}.
\end{gather}
Without the phase terms, the above matrix has a well known solution for a free particle 
with open boundary conditions. By modifying the well known open boundary solution with 
phase factors, we can find the eigenpairs of $H_{2,2}^V(k) = -H_{1,1}^V(k)$ and therefore 
obtain the eigenvectors of $V^\dag H(k) V$. The eigenpairs of $H_{2,2}^V(k) = -H_{1,1}^V(k)$ are
\begin{gather}
\label{}
u_{s} = \sqrt{\frac{2}{M+1}}
\begin{pmatrix}
\sin(\frac{s \pi}{M+1}) \\ e^{i\theta} \sin(\frac{2s \pi}{M+1}) \\ \sin(\frac{3s \pi}{M+1}) \\ e^{i\theta} \sin(\frac{4s \pi}{M+1}) \\\ \sin(\frac{5s \pi}{M+1})\\ \vdots 
\end{pmatrix}
\end{gather}
\begin{gather}
\label{}
\lambda_s = |x(k)| + 2 z \ \cos(\frac{s \pi}{M+1}).
\end{gather}
The eigenpairs of $V^\dag H(k) V$ are then 
$\lbrace \begin{pmatrix} u_{s} \\ 0 \end{pmatrix} \ , \ -\lambda_s\rbrace$ and $\lbrace \begin{pmatrix} 0 \\ u_{s} \end{pmatrix} \ , \ \lambda_s\rbrace$. 
We multiply these eigenvectors by $V$ to obtain the eigenvectors of $H(k)$. The eigenpairs 
of $H(k)$ are 
\begin{equation}
\label{}
w_s^- = \frac{1}{\sqrt{2}}
\begin{pmatrix}
-e^{i\theta} u_s^* \\ u_s
\end{pmatrix}
\ , \ E_s^- = -|x(k)| - 2 z \ \cos(\frac{s \pi}{M+1}) 
\end{equation}
\begin{equation}
\label{}
w_s^+ = \frac{1}{\sqrt{2}}
\begin{pmatrix}
e^{i\theta} u_s^* \\ u_s
\end{pmatrix}
\ , \ E_s^+ = |x(k)| + 2 z \ \cos(\frac{s \pi}{M+1}).
\end{equation}
So long as $|x(k)|>2z \cos(\frac{s \pi}{M+1})$ for all $s$ and $k$, the $w_s^-$ are always 
the eigenvectors below the Fermi level. This condition is equivalent to 
$0 \leq |z|\cos(\frac{\pi}{M+1})< |\delta|$. Then $Q(k) = I_{2M} - 2P(k) = I_{2M} - 2\sum_{s=1}^M w_s^- {w_s^-}^\dag$. 
Using the basic sum rules for sine waves, we have
\begin{equation}
Q(k) = 
\begin{pmatrix} 0_M & q(k) \\ q^\dag(k) & 0_M \end{pmatrix}
\end{equation}
with
\begin{equation}
\label{}
q(k) = \begin{pmatrix} e^{i\theta} & 0 & 0 & \dots \\ 0 & e^{-i\theta} & 0 & \dots \\ 0 & 0 & e^{i\theta} & \dots \\ \vdots & \vdots & \vdots & \ddots \end{pmatrix}.
\end{equation}
So the topological invariant is
\begin{eqnarray}
\label{Ifinal}
I &=& \frac{\im}{2 \pi} \mbox{tr} \int_{BZ}dk \ q^{-1}(k) \partial_k q(k) \nonumber \\ 
&=&
\begin{cases}
0, & \text{for } M \text{ even} \\
-\frac{1}{2 \pi}  \int_{BZ}dk \  \frac{\partial \theta}{\partial k}, & \text{for } M \text{ odd}\, .
\end{cases}
%
%
%
\end{eqnarray}
We know that the integral in the $M$ odd case is equal to $-2 \pi$ if and only if 
$\theta$ winds around the origin. That happens when $\delta>0$. This completes the proof.

\begin{figure}[h!]
\centering
\includegraphics[width=0.99\columnwidth]{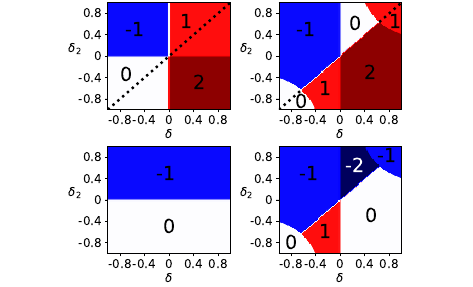}
\caption{Phase diagrams for the topological invariants $I$ and $I_3$ for the $M=3$ leg ladder. A dotted line is placed where the analytical result from the main text, Eq.~\eqref{winding_ana},
applies. The top diagrams are for $I$ and the bottom diagrams are for
$I_3$. Left panels: $ z=0.0$, Right panels: $z=0.9$.}
\label{Fig6}
\end{figure}
\label{AppC}
\setcounter{equation}{0}
\renewcommand{\thesection}{\text{C}}
\section{Additional Chiral Symmetries and Topological Invariants}
Here we give examples of additional chiral symmetries that can be defined
under certain conditions. Take, for example, the case of the $M=2$ SSH
ladder. Under the constraint $\delta_1=\delta_2$, we can define
another chiral symmetry operator $\hat{S}_2$. The transformation is
defined as
\small
\begin{gather}
\hat{S}_2 a_j^1 \hat{S}_2^{-1} = -i{a_j^2}^\dag \ , \  
\hat{S}_2 a_j^2 \hat{S}_2^{-1} = i{a_j^1}^\dag \nonumber \\
\hat{S}_2 b_j^1 \hat{S}_2^{-1} = i{b_j^2}^\dag \ , \ 
\hat{S}_2 b_j^2 \hat{S}_2^{-1} = -i{b_j^1}^\dag \\
\hat{S}_2 i \hat{S}_2^{-1} = -i.  \nonumber
\end{gather}
\normalsize
One can compute the topological invariant $I_2$ corresponding to the chiral symmetry $\hat{S}_2$. One can switch $j \rightarrow k_n$ to get the corresponding symmetry in momentum space. Then we can use $\hat{S} \psi_n \hat{S}^{-1} = \sum_{m}
(U_S)_{n,m}^* \psi_{m}^\dag$ to get the matrix $U_{S_2}(k)$. For this symmetry, we can work in the same basis as Eq.~\eqref{BlockUnitary} for $M=2$. In this basis,
\begin{equation}
U_{S_2}(k) = \begin{pmatrix} 
0 & 0  & 0 & i \\
0 & 0  & i & 0 \\
0 & -i  & 0 & 0 \\
-i & 0  & 0 & 0 
\end{pmatrix}.
\end{equation}
Then we can plug this directly into Eq.~(5) of the main text and solve numerically. We find that the invariant $I_2$ is zero for $-1 < \delta_1=\delta_2<1$ and $-1 < z<1$.

As another example, we consider the case of the $M=3$ SSH
ladder. Suppose we have the constraint $\delta_1=\delta_3$. Then
another chiral transformation is
\small
\begin{gather}
\hat{S}_3 a_j^1 \hat{S}_3^{-1} = {a_j^3}^\dag \ , \  
\hat{S}_3 a_j^2 \hat{S}_3^{-1} = -{a_j^2}^\dag \ , \
\hat{S}_3 a_j^3 \hat{S}_3^{-1} = {a_j^1}^\dag \nonumber \\
\hat{S}_3 b_j^1 \hat{S}_3^{-1} = -{b_j^3}^\dag \ , \  
\hat{S}_3 b_j^2 \hat{S}_3^{-1} = {b_j^2}^\dag \ , \
\hat{S}_3 b_j^3 \hat{S}_3^{-1} = -{b_j^1}^\dag  \\
\hat{S}_3 i \hat{S}_3^{-1} = -i. \nonumber
\end{gather}
\normalsize
Similarly, in the same basis as Eq.~\eqref{BlockUnitary} for $M=3$, we obtain
\begin{equation}
U_{S_3}(k) = \begin{pmatrix} 
0 & 0  & 1 & 0 & 0 & 0 \\
0 & 1  & 0 & 0 & 0 & 0 \\
1 & 0  & 0 & 0 & 0 & 0 \\
0 & 0  & 0 & 0 & 0 & -1 \\
0 & 0  & 0 & 0 & -1 & 0 \\
0 & 0  & 0 & -1 & 0 & 0 
\end{pmatrix}.
\end{equation}
Figure~\ref{Fig6} shows a comparison of the invariants $I$ from the main
text and $I_3$ defined here under the condition $\delta_1=\delta_3
\equiv \delta$. A dotted line is placed where the analytical result from the main text, Eq.~\eqref{winding_ana},
applies. The magnitude of $I_3$ indicates the number of topological
edge states protected by symmetry $\hat{S}_3$. Interestingly, for
$z=0.9$ shown in the right panels of Fig.~\ref{Fig6}, there is a
region where the invariant $I_3$ is equal to $-2$ while the invariant
$I$ is equal to zero. In this region, there are two topological edge
states protected by the symmetry $\hat{S}_3$.

\label{AppD}
\setcounter{equation}{0}
\renewcommand{\thesection}{\text{D}}
\section{Density Matrix Elements}
It is helpful to note
\begin{eqnarray}
\label{}
P^A A_1^\dag A_1 P^A &=& (A_1^\dag A_1 - A_1^\dag A_1 A_2^\dag A_2), \nonumber \\
P^A A_1^\dag A_2 P^A &=& A_1^\dag A_2
\end{eqnarray}
and similarly for the $B$ operators. Now we simply list a few important equations. The matrix elements not listed are similar.
\begin{eqnarray}
&&\langle P^B P^A P^A P^B\rangle=\langle P^A P^B \rangle = 
\langle A_1^\dag A_1 B_1^\dag B_1 \rangle \nonumber \\ 
&& +\langle A_1^\dag A_1 B_2^\dag B_2\rangle +\langle A_2^\dag A_2 B_1^\dag B_1\rangle+\langle A_2^\dag A_2 B_2^\dag B_2 \rangle \nonumber \\ 
&&-2\langle A_1^\dag A_1 B_1^\dag B_1 B_2^\dag B_2\rangle-2\langle A_2^\dag A_2 B_1^\dag B_1 B_2^\dag B_2\rangle \nonumber \\ 
&& -2\langle A_1^\dag A_1 A_2^\dag A_2 B_1^\dag B_1 \rangle-2\langle A_1^\dag A_1 A_2^\dag A_2 B_2^\dag B_2 \rangle \nonumber \\ 
&& +
4\langle A_1^\dag A_1 A_2^\dag A_2 B_1^\dag B_1 B_2^\dag B_2\rangle \qquad
\end{eqnarray}
\begin{eqnarray}
&& \langle \Psi_{1,1}|A_1^\dag B_1^\dag B_1 A_1|\Psi_{1,1}\rangle \\
&&= 
\frac{1}{\langle P^A P^B\rangle} ( \langle A_1^\dag A_1 B_1^\dag B_1\rangle -\langle A_1^\dag A_1 B_1^\dag B_1 B_2^\dag B_2\rangle  \nonumber \\ 
&& - \langle A_1^\dag A_1 A_2^\dag A_2B_1^\dag B_1\rangle +\langle A_1^\dag A_1 A_2^\dag A_2 B_1^\dag B_1 B_2^\dag B_2\rangle ) \nonumber
\end{eqnarray}
\begin{eqnarray}
&& \langle \Psi_{1,1}|A_1^\dag B_1^\dag B_2 A_1|\Psi_{1,1}\rangle \\
&& = 
\frac{1}{\langle P^A P^B\rangle}( \langle A_1^\dag A_1 B_1^\dag B_2\rangle- \langle A_1^\dag A_1 A_2^\dag A_2 B_1^\dag B_2\rangle ) \nonumber
\end{eqnarray}
\begin{equation}
\langle \Psi_{1,1}|A_1^\dag B_1^\dag B_2 A_2|\Psi_{1,1}\rangle  
= 
\frac{1}{\langle P^A P^B\rangle} \langle A_1^\dag A_2 B_1^\dag B_2\rangle
\end{equation}
\begin{eqnarray}
\langle \sigma_z^a \sigma_z^b\rangle_{1,1} = 
\frac{1}{\langle P^A P^B\rangle}(\langle A_1^\dag A_1 B_1^\dag B_1\rangle-\langle A_1^\dag A_1 B_2^\dag B_2\rangle \nonumber \\-\langle A_2^\dag A_2 B_1^\dag B_1\rangle+\langle A_2^\dag A_2 B_2^\dag B_2\rangle) \qquad
\end{eqnarray}
\begin{eqnarray}
\langle \sigma_z^a \sigma_x^b\rangle_{1,1} =
\frac{1}{\langle P^A P^B\rangle}(\langle A_1^\dag A_1 B_1^\dag B_2  \rangle+\langle A_1^\dag A_1 B_2^\dag B_1\rangle \nonumber \\-\langle A_2^\dag A_2 B_1^\dag B_2  \rangle-\langle A_2^\dag A_2 B_2^\dag B_1  \rangle) \qquad
\end{eqnarray}
\begin{eqnarray}
\langle \sigma_x^a \sigma_z^b\rangle_{1,1} = 
\frac{1}{\langle P^A P^B\rangle}(\langle B_1^\dag B_1 A_1^\dag A_2  \rangle+\langle B_1^\dag B_1 A_2^\dag A_1  \rangle \nonumber \\-\langle B_2^\dag B_2 A_1^\dag A_2  \rangle-\langle B_2^\dag B_2 A_2^\dag A_1  \rangle) \qquad
\end{eqnarray}
\begin{eqnarray}
\langle \sigma_x^a \sigma_x^b\rangle_{1,1} = 
\frac{1}{\langle P^A P^B\rangle}(\langle A_1^\dag A_2 B_1^\dag B_2  \rangle+\langle A_1^\dag A_2 B_2^\dag B_1  \rangle \nonumber \\+\langle A_2^\dag A_1 B_1^\dag B_2  \rangle+\langle A_2^\dag A_1 B_2^\dag B_1  \rangle) \qquad
\end{eqnarray}
\normalsize
\label{AppE}
\setcounter{equation}{0}
\renewcommand{\thesection}{\text{E}}
\begin{figure}[t!]
\centering
\includegraphics[width=0.99\columnwidth]{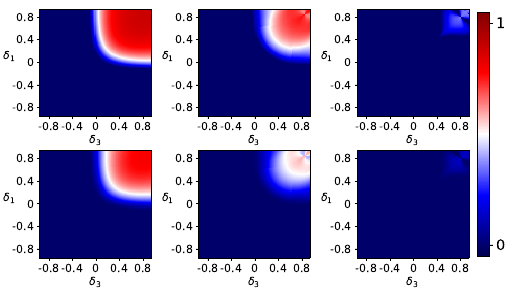}
\caption{Comparison of the entanglement phase diagram based on $E_{\textrm{neg}}[\rho_{AB}^{1,1}]$ (top row) and $E_{F}[\rho_{AB}^{1,1}]$ (bottom row)
in units of $\ln 2$ for a $3$-leg ladders with $L=16$ and $z=0.9$. Left to right:
$\delta_2=-0.75$; $\delta_2=-0.25$; $\delta_2=0.25$.}
\label{Fig7}
\end{figure}
\section{Entanglement of Formation}
Here we consider the entanglement of formation as an alternative
measure of bipartite entanglement of a mixed state. Given a density
matrix $\rho$, we define the entanglement of formation as
\cite{BennetFormation}
\begin{equation}
E_{F}(\rho) = \min \sum_i p_i E(|\psi_i\rangle)
\end{equation}
where E is the pure state von-Neumann entanglement and the
minimization is taken over all possible ensembles of the form $\rho =
\sum_i p_i |\psi_i \rangle \langle \psi_i|$.

One of the problems with the entanglement of formation is the
numerical difficulty of minimizing over all possible ensembles
\cite{Huang}. However, there is a method of obtaining the entanglement
of formation for any two qubit system \cite{Wooters}. First, define a
matrix $\tilde{\rho} = (\sigma_y \otimes \sigma_y) \rho^* (\sigma_y
\otimes \sigma_y)$ where $\rho^*$ is the complex conjugate of the
matrix $\rho$. Then define $R =
\sqrt{\sqrt{\rho}\tilde{\rho}\sqrt{\rho}}$.  and let $\lambda_i$
represent the eigenvalues of $R$ in decreasing order. The eigenvalues
$\lambda_i$ are also the square roots of the eigenvalues of $\rho
\tilde{\rho}$. The concurrence is then $C(\rho) = \max \lbrace 0,
\lambda_1-\lambda_2-\lambda_3-\lambda_4 \rbrace$. Define a variable $x
= \frac{1}{2} + \frac{1}{2}\sqrt{1-C^2(\rho)}$.  The entanglement of
formation is then $E_{F} = -x \ln(x)-(1-x)\ln(1-x)$. In
Fig.~\ref{Fig7}, results for the $M=3$ SSH ladder are compared to the
logarithmic negativity used in the main text. We can see that the
results are virtually the same.

\label{AppF}
\setcounter{equation}{0}
\renewcommand{\thesection}{\text{F}}
\section{Canonical/Grand Canonical Ensemble}
In this work, we used Wick's Theorem to calculate the n-point
correlation functions. Wick's Theorem, commonly used for
non-interacting pure states, also has a version for ensemble
calculations. Since the grand canonical ensemble has a factorizable
density matrix, Wick's Theorem for the grand canonical case is similar
to the pure state case \cite{Schonhammer}. Consider an operator
\begin{equation}
A_{k_1 , k_2 , \dots , k_m , \ell_1 , \ell_2 , \dots , \ell_m} = c^\dag_{k_1} c^\dag_{k_2} \dots c^\dag_{k_m} c_{\ell_m} \dots c_{\ell_2} c_{\ell_1}
\end{equation}
where none of the indices $k_i$ are the same, none of the indices $l_i$ are the same and the $c$ operators are in an arbitrary basis. Define an $m \times m$ matrix $(\langle c^\dag c \rangle)_{i,j} = \langle c^\dag_{k_i} c_{l_j}\rangle$. Then for pure states and in the grand canonical ensemble 
\begin{equation}
\label{WickGrandCanonical1}
\langle A_{k_1 , k_2 , \dots , k_m , \ell_1 , \ell_2 , \dots , \ell_m} \rangle = \text{det} \langle c^\dag c \rangle.
\end{equation}
For the grand canonical ensemble, we can use the Fermi-Dirac distribution for the occupation number $n_i = c^\dag_i c_i$ expectation value in the diagonal basis \cite{ShapourianRyu}
\begin{equation}
\label{WickGrandCanonical2}
\langle n_i \rangle = \frac{1}{1+ e^{\beta(\epsilon_i-\mu)}}.
\end{equation}
In the grand canonical ensemble, we set the chemical potential $\mu=0$
in order to preserve particle-hole symmetry. For the canonical
ensemble, \eqref{WickGrandCanonical1} and \eqref{WickGrandCanonical2}
are not satisfied. In the canonical ensemble, multi-point correlators
can be written in terms of two point correlators in the diagonal basis
as
\begin{figure}[t!]
\centering
\includegraphics[width=0.99\columnwidth]{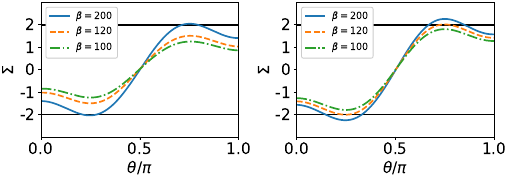}
\caption{$\Sigma$, as defined in Eq.~(9) of the main text, for a 3-leg ladder with $L=8$, $\delta_1=0.29$, $\delta_2=-0.99$, $\delta_3=0.99$ and $z=0.95$ for various inverse temperatures $\beta=1/T$. Left panel: Grand canonical ensemble with chemical potential $\mu=0$. Right panel: Canonical ensemble.}
\label{Fig8}
\end{figure}

\begin{equation}
\langle n_{\ell_1}   n_{\ell_2}  \dots n_{\ell_m} \rangle = \sum_{i=1}^m \langle n_{\ell_i} \rangle \prod_{j (\neq i)}^m \frac{e^{\beta \epsilon_{\ell_i}}}{e^{\beta \epsilon_{\ell_i}}-e^{\beta \epsilon_{\ell_j}}}.
\end{equation}
Note that the equation above is only valid if all of the energy levels
$\epsilon_{\ell_i}$ are nondegenerate, which we have verified numerically for the results in this section. The distribution $\langle n_i
\rangle$ itself has to be calculated using a numerically stable
recursion relation \cite{Borrmann}. Figure~\ref{Fig8} shows the breaking
of the Bell's inequality for $\Sigma$ as defined in Eq.~(9) of the
main text. The grand canonical ensemble breaks the inequality for
$\beta$ as low as $200$ and the canonical ensemble breaks the
inequality for $\beta$ as low as $120$. For a cold atomic gas
experiment---neglecting particle loss---the question whether the
canonical or the grand-canonical ensemble are more appropriate depends
on how well-controlled the particle number for each realization is. As
expected, stronger quantum correlations are preserved to higher
temperatures in the canonical ensemble.

\end{document}